\documentclass[useAMS,usenatbib,a4paper]{mn2e}
\usepackage{graphicx}
\usepackage{deluxetable1}
\usepackage{amsmath}
\usepackage{amssymb}

\title[GRB 090426] {GRB 090426: The Environment of a Rest-Frame 0.35-second Gamma-Ray Burst at
Redshift z=2.609}
\author[E. M. Levesque et al.]{Emily M. Levesque$^{1,2}$, Joshua S. Bloom$^{3}$, Nathaniel R. Butler$^{3}$\thanks{GLAST/Einstein Fellow}, Daniel A. Perley$^{3}$,
\newauthor S. Bradley Cenko$^{3}$, J. Xavier Prochaska$^{4}$, Lisa J. Kewley$^{1}$, Andrew Bunker$^{5}$, 
\newauthor Hsiao-Wen Chen$^{6}$,Ryan Chornock$^{3}$, Alexei V. Filippenko$^{3}$, Karl Glazebrook$^{7}$, 
\newauthor Sebastian Lopez$^{8}$, Joseph Masiero$^{1}$, Maryam Modjaz$^{3}$\thanks{Miller Fellow.}, Adam N. Morgan$^{3}$,
\newauthor and Dovi Poznanski$^{3}$\\
$^{1}$Institute for Astronomy, University of Hawaii, 2680 Woodlawn Dr., Honolulu, HI, 96822, USA.\\
$^{2}$Smithsonian Astrophysical Observatory, 60 Garden St., Cambridge, MA, 02139, USA.\\
$^{3}$Department of Astronomy, 601 Campbell Hall, University of California, Berkeley, CA, 94720-3411, USA.\\
$^{4}$Department of Astronomy and Astrophysics, UCO/Lick Observatory, University of California, 1156 High Street, Santa Cruz, CA, 95064, USA.\\
$^{5}$Department of Astrophysics, Oxford University, Keble Road, Oxford, OX1 3RH, United Kingdom.\\
$^{6}$Department of Astronomy \& Astrophysics, University of Chicago, Chicago, IL 60637, USA.\\
$^{7}$Centre for Astrophysics and Supercomputing, Swinburne University of Technology, P.O. Box 218 Hawthorn, VIC 3122, Australia.\\
$^{8}$Departamento de Astronomia, Universidad de Chile, Casilla 36-D, Santiago, Chile.}
\begin{document}

\date{Accepted 1988 December 15. Received 1988 December 14; in original form 1988 October 11}

\pagerange{\pageref{firstpage}--\pageref{lastpage}} \pubyear{2002}


 

\date{Released 2009 Xxxxx XX}

\pagerange{\pageref{firstpage}--\pageref{lastpage}} \pubyear{2009}

\def\LaTeX{L\kern-.36em\raise.3ex\hbox{a}\kern-.15em
    T\kern-.1667em\lower.7ex\hbox{E}\kern-.125emX}

\newtheorem{theorem}{Theorem}[section]

\newcommand{\citepeg}[1]{\citep[{e.g.,}][]{#1}}
\newcommand{\citepcf}[1]{\citep[{see}\phantom{}][]{#1}}
\newcommand{\rha}[0]{\rightarrow}
\def\etal{{\sl et al.}}
\def\lsim{\hbox{ \rlap{\raise 0.425ex\hbox{$<$}}\lower 0.65ex\hbox{$\sim$}}}
\def\gsim{\hbox{ \rlap{\raise 0.425ex\hbox{$>$}}\lower 0.65ex\hbox{$\sim$}}}
\def\arcmin{\hbox{$^\prime$}}
\def\arcsec{\hbox{$^{\prime\prime}$}}
\def\arcdeg{\mbox{$^\circ$}}
\def\fd{\hbox{$~\!\!^{\rm d}$}}
\def\fh{\hbox{$~\!\!^{\rm h}$}}
\def\fm{\hbox{$~\!\!^{\rm m}$}}
\def\fs{\hbox{$~\!\!^{\rm s}$}}
\def\ale{\mathrel{\hbox{\rlap{\hbox{\lower4pt\hbox{$\sim$}}}\hbox{$<$}}}}
\def\age{\mathrel{\hbox{\rlap{\hbox{\lower4pt\hbox{$\sim$}}}\hbox{$>$}}}}
\def\msyr{\hbox{M$_\odot$ yr$^{-1}$}}

\label{firstpage}

\maketitle

\begin{abstract}
We present the discovery of an absorption-line redshift of $z = 2.609$ for GRB 090426, establishing the first firm lower limit to a redshift for a gamma-ray burst with an observed duration of
$<2$~s. With a rest-frame burst duration of $T_{90z} = 0.35$ s and a detailed examination of the peak energy of the event, we suggest that this is likely (at $>$90\% confidence)
a member of the short/hard phenomenological class of GRBs. From analysis of the optical-afterglow spectrum we find
that the burst originated along a very low HI column density
sightline, with $N_{\rm HI} < 3.2 \times 10^{19}$ cm$^{-2}$. Our GRB 090426 afterglow spectrum also appears to have
weaker low-ionisation absorption (Si II, C II) than $\sim$95\% of
previous afterglow spectra. Finally, we
also report the discovery of a blue, very luminous, star-forming putative host galaxy
($\sim2L_*$) at a small angular offset from the location of the
optical afterglow. We consider the implications of this unique GRB in
the context of burst duration classification and our understanding of
GRB progenitor scenarios.
\end{abstract}


\section{Introduction}
The early emergence of a two-class bifurcation in the high-energy
properties of gamma-ray bursts \citep[GRBs;][]{1993ApJ...413L.101K}
gave rise to the supposition that two distinct ``progenitors'' could
be responsible for the lion's share of such events (e.g.,
\citealt{2004IJMPA..19.2385Z}). While observations directly link
long-duration soft-spectra GRBs (LSBs) to the death of young massive
stars (\citealt{stanek03,hjorth03}; see \citealt{Woosley:2006p189} for
a review), less-strong circumstantial evidence (based on physical
associations with more evolved galaxies) suggests that at least some
fraction of short-duration hard-spectra GRBs (SHBs) are due to older
progenitors
\citep{2006ApJ...638..354B,ffp+05,bpc+05,2005ApJ...630L.117H,2006ApJ...642..989P,2006AIPC..836..473B}\footnote{In this paper, we will use the term ``SHB" to denote this phenomenological \textit{class} of gamma-ray bursts as first identified in the BATSE sample.  This is distinct from a simple cut on duration or hardness, since the populations are known to overlap to varying degrees with different instruments. Furthermore, identification with a class does not necessarily imply identification with a particular progenitor, even though most LSBs have been associated with massive stars and a few SHBs have been associated with old populations.}. Whether
SHBs are due to the coalescence of two neutron stars, other compact
degenerate binaries, some combination of these models, or something else entirely is an open question
(cf.\ \citealt{2007NJPh....9...17L}).

Several lines of evidence now suggest that the true progenitor
diversity does not map with one-to-one correspondence to the two-class
phenomenological landscape\footnote{See \citet{2008AIPC.1000...11B} and \citet{Zhang07} for a discussion of GRB classification, both physical and phenomenological.}. In particular, there appear to be many
more than just two progenitors.  For example, a small fraction of SHBs
probably originate from massive flaring activity of extragalactic
magnetars \citep[highly magnetised neutron stars;][]
{2008ApJ...681.1419A,2008ChJAS...8..202H,2009MNRAS.395.1515C}. Either
similar magnetar activity (or perhaps a flaring accretion-powered
system) from objects in our own Galaxy may occasionally create LSBs as
well \citep{2008ApJ...678.1127K,2008Natur.455..506C}.  Classification
of individual events even among the two well-established cosmological
groups has proven extremely difficult, not only in the overlap region of the duration-hardness diagram where the population distributions merge (at around 1$-$2 seconds for {\it Swift}/BAT) but even for much longer-duration events.  Indeed, some
of the same SHBs which have been used to link this phenomenological
class to older, evolved galaxies actually have observed total
durations (as measured by $T_{90}$, the interval over which 90\% of
the burst counts are observed) of over 100~s due to a component
of extended emission (EE) that follows the initial spike.  At least
two LSBs at low redshift, GRB 060505 and GRB 060614, were not accompanied by observable supernovae
despite intense follow-up campaigns
\citep{fwt+06,2006Natur.444.1044G}, and it is still debated whether
these events group most naturally with short-duration events,
long-duration events, or a new class entirely (e.g. \citealt{Jakobsson07,Lu08,Xu09,Levesque07,Thone08}, and others).  Most recently, the two
highest redshift GRBs detected to date, at $z$ = 6.7 and $z$ = 8.2
\citep{Greiner+2009,Salvaterra+2009,Tanvir+2009,Zhang09}, were observed to
have rest-frame durations of $<2$~s, yet few have argued that these events did not arise from massive stars.

To date, the strongest evidence that many SHBs and LSBs arise from a different progenitor population comes from
analysis of their respective host galaxy associations. The
host galaxies of long-duration GRBs have, universally, been observed
to have blue colours, sub-solar ISM metallicities, and strong emission features associated with high
specific star-formation rates \citep{Savaglio+2009,Stanek06,Modjaz08,Berger09}.  The burst
position, when well-constrained, is nearly always at small
offset \citep{Bloom+2002} and typically traces the brightest regions of
the host galaxy \citep{Fruchter+2006}, which itself is typically blue
and morphologically disturbed \citep{Wainwright+2007}. \citet{Fruchter+2006} also find that LGRB host galaxies have lower luminosities on average compared to the galaxy population probed by surveys at similar redshifts.  In contrast,
the host galaxies of short-duration GRBs to date have been observed to
be much more heterogeneous, including both star-forming and non
star-forming hosts.  Afterglow offsets range from negligible to many
times the half-light radius of the putative
host \citep{2006ApJ...642..989P,Bloom+2007,Berger07,2008MNRAS.385L..10T}.

It is against this backdrop that GRB 090426 enters the scene.  With an
observed duration of $T_{90} = 1.28 \pm 0.09$ s (\S 2), based on its
observer-frame duration alone it groups more closely with the SHB
class, an identification that is further bolstered when its high
redshift of $z = 2.609$ (\S \ref{sec:spec}) is considered, implying a
rest-frame duration of only 0.35~s that is unambiguously within
the range of classical SHBs.  By contrast, the most convincing host associations for SHBs
are at $z < 1$ (although see \citealt{2007ApJ...664.1000B} and a discussion of the potentially high-redshift SHB GRB 060121; \citealt{2006ApJ...648L..83D,2006ApJ...648L...9L}). 

Irrespective of the phenomenological classification or the physical origin of this event, we report on an optical spectrum of the afterglow --- the first ever reported for an event with an observed duration
of $<2$~s --- which shows evidence of an environment quite unlike that
of most (but not unprecedented among) GRBs of long duration with spetroscopic observations. We also
present the results from our campaign of late-time imaging and
spectroscopy, which identify the highly UV-luminous host galaxy of
this event.  All of the values and conclusions in this paper are
consistent with our GCN Circulars\footnote{The GCN system {{\tt http://gcn.gsfc.nasa.gov/}} is managed and operated by Scott Barthelmy.} but should be considered to
supersede our previous work on this event.
 
\section{The Discovery and Classification of GRB 090426}
At 12:48:47 on 2009 April 26\footnote{UT dates are used throughout this
paper.}, the Burst Alert Telescope (BAT; \citealt{Barthelmy05}) onboard the NASA {\it Swift}
satellite \citep{Gehrels04} was triggered on GRB 090426. The X-Ray
Telescope (XRT; \citealt{Burrows05}) began observing the field at 84.6~s after the trigger,
and the ultraviolet/optical telescope (UVOT; \citealt{Roming05}) followed at 89~s after
the trigger. UVOT detected a candidate optical afterglow at $\alpha =
12\fh36\fm18\fs.07$, $\delta = +32^\circ59\arcmin09\arcsec.6$ (J2000),
which was reported by \citet{cummings09} 13.8 min after the burst
trigger. The optical counterpart at these coordinates was also
confirmed 43.5 min after the burst by \citet{xin09} based on
observations obtained 76~s after the burst with the Teramo-Normale
Telescope. The Sloan Digital Sky Survey (SDSS) shows no object near
the position of the afterglow; the closest object is a faint and
extended source at $\alpha = 12\fh36\fm19\fs.49$, $\delta =
+32^\circ59\arcmin05\arcsec.5$ (J2000), 18\arcsec\ away from the
optical afterglow with a photometric redshift of $z \sim 0.3$
\citep{davanzo09}. As detailed in \S 4.1, we obtained a spectrum of
the afterglow 1.1 hr after trigger, independently discovering the
optical afterglow by inspection of the guider and acquisition
frames and determining an absorption redshift of $z = 2.609$ \citep{levesque09}. Later, \citet{thone09} confirmed this afterglow detection and
redshift with a Very Large Telescope (VLT) spectrum observed 12.3~hr
after the trigger.

\subsection{The Short-Duration/Hard-Spectrum Bifurcation in {\it Swift}}
\label{sec:bi}

\begin{figure*}
\includegraphics{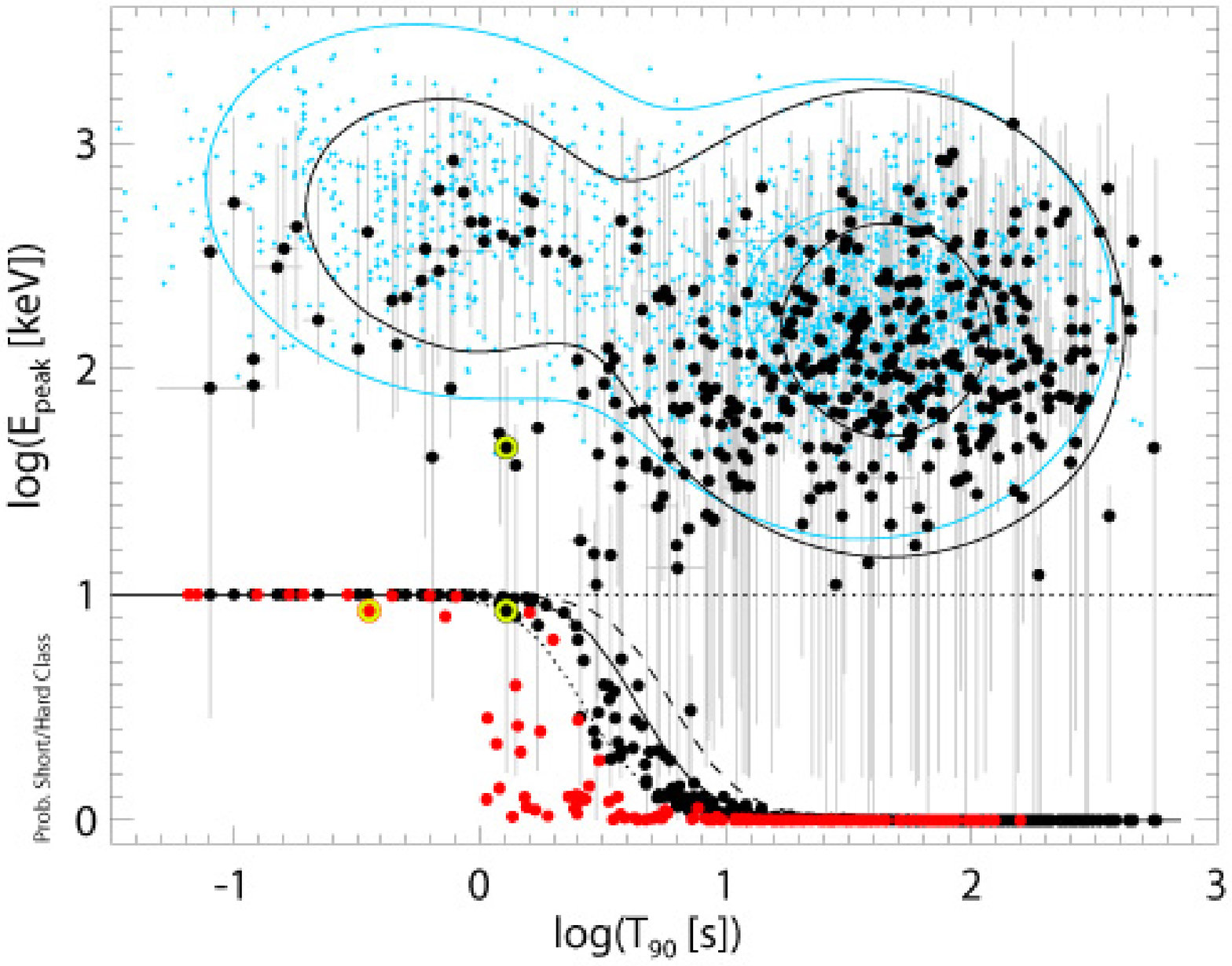}
\caption{{\it Top}: The $T_{90}$ durations and $E_{\rm peak}$ values
  for 398 {\it Swift} GRBs.  In the background, we plot (small blue circles)
  1728 $T_{90}$ and $E_{\rm peak}$ values for historic GRBs detected
  by the BATSE experiment.  Overplotted is the best-fit double-Gaussian 
  model to the BATSE data (blue; 50\% peak probability and
  5\% peak probability contours).  The black curves show the relative
  distortion expected for these distributions appearing in {\it Swift},
  given the relative satellite detection efficiencies (see Band 2006).
  Because of its increased relative capacity to trigger on long/soft
  events, there are relatively fewer (by a factor of $\sim$3) short/hard
  events compared to long/soft events in the {\it Swift} sample as
  compared to the BATSE sample, making classification given the {\it Swift}
  data alone difficult if not impossible.  GRB~090426 is highlighted
  and circled in yellow. {\it Bottom}: The ratio of Gaussians defines
  the probability that a given burst will belong to the short/hard
  class. Red circles give the host-frame $T_{90z} = T_{90}/(1+z)$
  values for 138 GRBs with measured spectroscopic redshifts. The solid
  black curve shows the projection onto the abscissa of the solid
  distribution in the top panel. Dashed and dotted curves are also shown to
  display the $E_{\rm peak}$ dependence of the curve at two $E_{\rm
    peak}$ values ($E_{\rm peak} = 1$ MeV and $E_{\rm peak} =20$ keV,
  respectively). GRB 090426 $T_{90}$ and $T_{90z}$ are circled in
  yellow.}
\label{fig:t90}
\end{figure*}

The question of which phenomenological classification to ascribe to
GRB 090426 is obviously an important one. To do so, we examine the
hardness--duration distribution of {\it Swift} and BATSE (from which
the original classification scheme was derived). Here, we measure
hardness by fitting a \citet{1993ApJ...413..281B} model to the BAT
spectrum and extracting the best-fit $\nu F_{\nu}$ peak energy $E_{\rm
  peak}$.  Figure \ref{fig:t90} (top panel) displays the durations and
hardnesses for 398 {\it Swift} GRBs detected by {\it Swift} between
December 2004 and April 2009.  Plotted in the background are
hardnesses and durations for 1728 historic GRBs detected by the BATSE
experiment, taken from the current BATSE
catalogue\footnote{http://www.batse.msfc.nasa.gov/batse/grb/catalog/current}.

Our methodology for determining $E_{\rm peak}$ (and $T_{90}$) from the
BAT data is described in detail by \citet{2007ApJ...671..656B}.  The
procedure involves prior assumptions on the
\citet{1993ApJ...413..281B} model parameters in order to overcome
challenges associated with fitting this broad-band spectral model to
data covering only the BAT 15--150 keV spectral band.  Large error
bars on $E_{\rm peak}$ necessarily result, as shown in the figure.  To
estimate $E_{\rm peak}$ for the largest possible number of BATSE GRBs,
we determine a relationship between the BATSE catalogue hardness ratio
(HR) and the measured $E_{\rm peak}$ from \citet{2006ApJS..166..298K}
for 325 GRBs in common.  We find $E_{\rm peak} \approx 80({\rm
  HR}_{3412})^{0.69}$ keV, with a scatter of 0.3 dex.  Here, HR$_{3412}$
is the hardness ratio of fluences in BATSE bands $3+4$ over $1+2$.

It has been noted previously (e.g.,
\citealt{2006HEAD....9.1370C,2008A&A...484..293Z}) that there is only weak
evidence for a distinct short-duration class in the {\it Swift} sample
considered alone.  \citet{2006ApJ...644..378B} suggests that the
discrepancy arises primarily as a result of the {\it Swift} increased
sensitivity (relative to BATSE) to long GRBs, which tends to make
detected short-duration GRBs a factor $\sim 3$ less common in {\it Swift}
relative to BATSE.  We explore this possibility in detail here by
correcting the observed {\it Swift} number distributions for sensitivity.
This is accomplished using the sensitivity curves as a function of
$E_{\rm peak}$ and $T_{90}$ duration from Figures 3 and 4 of Band
(2006).  The curves we use assume an exponential burst light curve and a
\citet{1993ApJ...413..281B} model with $\alpha=-1$ and $\beta=-2$.

We first fit a double elliptical Gaussian model to the observed
distributions in $T_{90}$ and $E_{\rm peak}$ from BATSE.  Assuming a
Gaussian shape entails making the fewest assumptions on the true
underlying distributions, because a Gaussian is the maximum-entropy
distribution in the case of known mean and variance
\citep[e.g.,][]{2005blda.book.....G}.  We employ a Markov Chain Monte
Carlo algorithm based on the data augmentation algorithm in
\citet{2001ApJ...548..224V} to propagate errors and marginalise over
the thirteen Gaussian parameters defining the best-fit, two-class model
shown in blue in Figure \ref{fig:t90} (top panel).  We begin by
stochastically dividing the BATSE observations between classes, given
an initial guess for the Gaussian parameters and also samples for the
$T_{90}$ and $E_{\rm peak}$ values from their respective best-fit
distributions (assumed Gaussian).  With this division in place, we
find the best-fit Gaussian parameters again and draw samples for each
from the posterior distribution using the Gibbs sampling technique
(e.g. \citealt{2001ApJ...548..224V}).  The process is iterated,
allowing us to determine $10^3$ samples for each parameter after
dropping 100 samples (``burn in'') to allow the chain to converge.

Next, we scale the best-fit double-class model by the relative
sensitivity curve to obtain the contours in black, which are those
expected for {\it Swift}.  To do the scaling, we must assume a
relation for the number of bursts gained (lost) as the sensitivity is
decreased (increased).  We assume that the number scales as the
relative sensitivity to some power $\eta$.  This $\eta$ is the slope
of the cumulative number density versus flux relation (i.e., the log
$N$--log $S$ relation, as in \citealt{2000ApJS..126...19P}).  We take
$\eta=-0.75$.  Our resulting black curves do appear to better match
the {\it Swift} $T_{90}$ and $E_{\rm peak}$ distributions (Figure
\ref{fig:t90}, top panel).

To quantitatively judge the validity of
the corrected model, we determine the rate increase/decrease factor
for each GRB as a function of $E_{\rm peak}$ and $T_{90}$ in the {\it
  Swift} sample and generate from these a corrected $T_{90}$
histogram.  The Kolmogorov-Smirnov (KS) test probability that the
uncorrected distributions from {\it Swift} and BATSE are the same is
$10^{-6}$; however, with application of the relative sensitivity
function to adjust the rate for each event, the KS-test probability is
only 2.7\%.  Hence, the distributions become only marginally
inconsistent with no ad hoc changes.  More precise tweaking, which would
utilise the exact spectral/temporal properties and a detailed
simulation of the {\it Swift} trigger algorithm, would likely improve
the consistency, although this analysis would be very challenging to
conduct and is beyond the scope of this work.

\subsection{Classification by $T_{90}$ and $E_{\rm peak}$ for GRB~090426}

We download the raw, unfiltered {\it Swift} BAT data for GRB~090426
from the 
{\it Swift}~Archive\footnote{ftp://legacy.gsfc.nasa.gov/swift/data.}.  
Our reduction of these data to science-quality light curves and spectra
using standard {\it Swift} tools are detailed by \citet{2007ApJ...671..656B}.  We
employ calibration files from the 2008-10-26 BAT release.  The BAT
signal in the 15--350 keV band consists of a single, narrow emission
spike of duration $T_{90} = 1.28 \pm 0.09$~s ($T_{50} = 0.48 \pm
0.06$~s).  The spectrum in the 15--150 keV band is well modelled
($\chi^2/\nu=54.18/45$) as a single power law with photon index
$\beta=-2.02^{+0.25}_{-0.28}$ and an energy fluence (15--350 keV) of
$2.5^{+0.4}_{-0.3} \times 10^{-7}$ erg cm$^{-2}$.  Using our Bayesian
methodology \citep{2007ApJ...671..656B} to extrapolate to an
approximately bolometric energy release in the 1--10$^4$ keV band
(source frame), we find $E_{\rm iso} = 4.2^{+5.9}_{-0.4} \times
10^{51}$ erg, with a $\nu F_{\nu}$ spectral peak energy of $E_{\rm
  peak} = 45^{+57}_{-43}$ keV (observer frame).

Above, we map a 2-class model (Gaussian $G_1$ for the short/hard class
and Gaussian $G_2$ for the long/soft class) justified based on comparing BATSE
data to the {\it Swift} sample.  We will now apply this classification to
{\it Swift} GRBs and to GRB~090426 in particular.  Importantly, the precise
factor dictating the relative {\it Swift}/BATSE number distribution at a
given value of $E_{\rm peak}$ and $T_{90}$ --- which we derive
approximately above as arising solely from variations in the satellite
sensitivities --- does not enter into this calculation.  We only need
to know the ratio of the probabilities, which means the factor drops
out of the relative classification calculation.  In principle, the
relative rate factor could also depend on variations with redshift of
the intrinsic source populations (ignored above) at fixed values
$E_{\rm peak}$ and $T_{90}$, but we make the simplifying assumption
here that this can be ignored.

It is important to demonstrate that our derived $E_{\rm peak}$ and
$T_{90}$ values from {\it Swift} are sufficiently similar to those derived
from BATSE.  Because we have defined $E_{\rm peak}$ in a similar fashion
for both experiments, this then primarily becomes an issue of
comparing $T_{90}$ values derived in different bandpasses, whereas we
know GRB spectral evolution tends to make a given event appear longer
when measured in a lower energy bandpass (e.g.,
\citealt{1995ApJ...448L.101F}).  Fortunately, the {\it Swift} (15--350 keV)
and BATSE (50--300 keV) bandpasses are similar, and we can directly
measure any biases in {\it Swift} $T_{90}$ values calculated in the 15--350
keV band as opposed to the 50--300 keV band.  Considering 411 {\it Swift}
GRBs, we find that the median decrease in $T_{90}$ when considering
the 50--300 keV band instead of the 15--350 keV band is only 3.8\%.
The decrease is $<30$\% for 90\% of the sample, which is typically
($>75$\% of the time) contained within our $1\sigma$ error bar on
$T_{90}$.  Therefore, we expect systematic variations in $T_{90}$
with energy band to not affect our classification.

Using the Markov Chain derived for the model division as a function of
$T_{90}$ and $E_{\rm peak}$, we can directly determine the
probabilistic class association for {\it Swift} GRBs taking into account
errors in $T_{90}$ and $E_{\rm peak}$.  Using a Markov Chain for this
purpose effectively treats the error bars on all quantities and allows
us to marginalise over the parameters describing the 2-class model.
The marginalisation is important, because there is strong
overlap in the observed BATSE distributions which translates to
uncertainty in the Gaussian model parameters defining our BATSE
classification.  To classify the {\it Swift} GRBs, we sample $10^3$ values
for $T_{90}$ and $E_{\rm peak}$ from the distributions in
\citet{2007ApJ...671..656B}. Each of these samples is used to evaluate
one of the $G_1/(G_1+G_2)$ draws above.  The class probability, which
is the Bayesian evidence in favour of the membership in the SHB class
as compared to the LSB class, is calculated as the median of the
$G_1/(G_1+G_2)$ samples.

The ratio of Gaussians $G_1/(G_1+G_2)$, evaluated at a particular
value for $T_{90}$ and $E_{\rm peak}$, defines the probability that a
given burst will belong to the SHB class under our assumptions.  In the bottom panel of
Figure \ref{fig:t90}, the solid black curve shows the projection onto
the abscissa of the solid distribution in the top panel.  This curve
is the classification marginalised over $E_{\rm peak}$.  Note that
because we have assumed a sensitivity correction as a function of
$T_{90}$ and $E_{\rm peak}$ at each value of $T_{90}$ and $E_{\rm
  peak}$, the correction drops out of the ratio $G_1/(G_1+G_2)$, and
we can apply the BATSE $G_1/(G_1+G_2)$ model to {\it Swift} without needing
to account for relative sensitivity.  Dashed curves are also shown to
display the $E_{\rm peak}$ dependence of the curve at two $E_{\rm
  peak}$ values ($E_{\rm peak} = 1$ MeV and $E_{\rm peak} =20$ keV,
respectively).

In black circles in Figure \ref{fig:t90} (bottom panel), we display the probability for each Swift burst plotted in the top panel. We also plot as red circles the host-frame $T_{90z} = T_{90}/(1+z)$ values for 138 GRBs with measured spectroscopic redshifts.  Considering the range of observed $E_{\rm peak}$ values, we find that a GRB is short/hard at $>90$\% confidence if $T_{90}<2.2$ s, or $T_{90z}<0.8$ s.  These limits can be used in future studies to select {\it Swift} GRBs belonging with high confidence to the short/hard class. Note that our spectroscopic redshift of $z= 2.609$ for GRB 090426 enables us to derive a rest-frame duration $T_{90z}$ of 0.35 $\pm$ 0.03 sec.

We find the probability that GRB 090426, highlighted and circled in yellow in Figure \ref{fig:t90}, belongs to the short/hard class is 92.8\%. Even so, we must stress that such a high confidence indication could occur by chance given a sufficient number of detected SHBs.  For $\sim400$ {\it Swift} LSBs detected to date, the chance probability of detecting one or more long/soft GRBs with durations short enough and/or hardness high enough to appear short/hard with such high confidence in our scheme is $>90$\%. This marks a fundamental shortcoming in the classification by high-energy properties alone, where the parameter distributions suffer broad overlap, motivating further investigation into the afterglow and host properties. It is possible that additional high energy indicators (e.g., a ``lag'' consistent with zero, \citealt{2009GCN..9272....1U}) may be useful for classification, but we do not investigate these here. 

\section{Energetics and Afterglow}

As mentioned above, from our Bayesian model of the burst parameters,
we calculate an isotropic energy release of $4.2 \times 10^{51}$ erg.  Until very recently, this value of $E_{\rm
  iso}$ would be considered exceptionally large for a SHB.  Indeed, a low
$E_{\rm iso}$ is naturally expected from most SHB models
\citep{2001ApJ...561L.171P}, which are typically assumed to collimate
their ejecta less efficiently than the collapsar model \citep{Berger07,Nakar07}.  However, the
recent GRB 090510 (spatially associated with an emission-line galaxy) had a very large $E_{\rm iso} = 3.8 \times 10^{52}$
erg \citep{Rau09}, suggesting that short-duration bursts are indeed
capable of arising from very energetic (and/or tightly collimated) explosions as well, and are probably
visible substantially beyond $z = 1$, even if they are not expected to
be common.

Combining the (limited) set of observations of the afterglow of GRB 090426 from published circulars\footnote{See http://gcn.gsfc.nasa.gov/gcn3\_archive.html}, we find the optical light curve is well described by a single power-law decay with index $\alpha_{O} \approx 0.8$ from $t \lesssim 100$s until $t \gtrsim 4 \times 10^{4}$s.  This is similar to the inferred X-ray decay index, $\alpha_{X} \approx 0.9$, from the online compilation of N.R.B.\footnote{http://astro.berkeley.edu/$\sim$nat/swift; see \citet{2007ApJ...671..656B} for details.}.  Combined with the derived optical to X-ray spectral index for this interval, $\beta_{OX} \approx 0.9$, and assuming standard synchrotron afterglow theory (e.g., \citealt{spn98}), these results suggest an afterglow with a shallow electron index ($p \approx 1.8$) and a cooling frequency $\nu_{c}$ below the optical bandpass.  If this is indeed the case, we can use the limit on the cooling frequency to constrain the parsec-scale circumburst density.  Assuming a constant density medium, the cooling frequency falling below the optical requires (e.g., \citealt{gs02}):
\begin{equation}
n \gtrsim 0.05 \epsilon_{B}^{-3/2} E_{\mathrm{KE,}52}^{-1/2},
\end{equation}              
where $n$ is the circumburst density (cm$^{-3}$), $\epsilon_{B}$ is the fraction of the shock energy partitioned to the magnetic field, and $E_{\mathrm{KE,}52}$ is the kinetic energy of the outgoing blastwave ($10^{52}$\,erg).  Given a maximal $\epsilon_{B}$ of 1/3 at equipartition, and with $E_{\mathrm{KE}} \lesssim 10 E_{\gamma} \approx 10^{53}$\,erg, we derive a lower limit of $n \gtrsim 0.1$\,cm$^{-3}$.  A similar result can be derived for the case of a wind-like circumburst medium (e.g., \citealt{cl99}): $A_{*} \gtrsim 0.01$ (where $\rho = 5 \times 10^{11} A_{*}$\,g\,cm$^{-1}$, chosen to correspond to a mass loss rate of $\dot{M} = 10^{-5}$ M$_{\odot}$\,yr$^{-1}$ and a wind speed of $v_{w} = 1000$\,km\,s$^{-1}$).  While we caution that this result is based on a relatively sparsely sampled optical light curve, it is clear that presence of a relatively bright and slowly fading optical and X-ray afterglow distinguish GRB\,090426 from the extremely low-density circumburst environments inferred for the short GRB\,080503 ($n \lesssim 5 \times 10^{-6}$\,cm$^{-3}$; \citealt{pmg+09}) or the long GRB\,050911 \citep{pkl+06}.     

\section{Optical Afterglow Spectrophotometry}
\label{sec:spec}
\subsection{Observations and Reductions}

We obtained an optical spectrum of the afterglow of GRB 090426 using
the Keck Low-Resolution Imaging Spectrometer (LRIS; \citealt{occ+95})
at 13:55 on 2009 April 26, $\sim$1.1 hr after the BAT trigger. The
observations were conducted in photometric conditions. We obtained two
300~s exposures on the LRIS blue side using the long 1\arcsec\ slit
mask, the 680 dichroic, and the 300/5000 grism. We observed internal flatfield lamps as well
as spectra of Hg, Ne, Ar, Cd, and Zn comparison lamps to be used for
wavelength calibration. We also obtained a 60~s spectrum of the
spectrophotometric standard HZ 43. The observations of the GRB 090426
afterglow were conducted at a high airmass of 3.05; HZ 43
was observed at an airmass of 3.60.

\begin{figure*}
\includegraphics{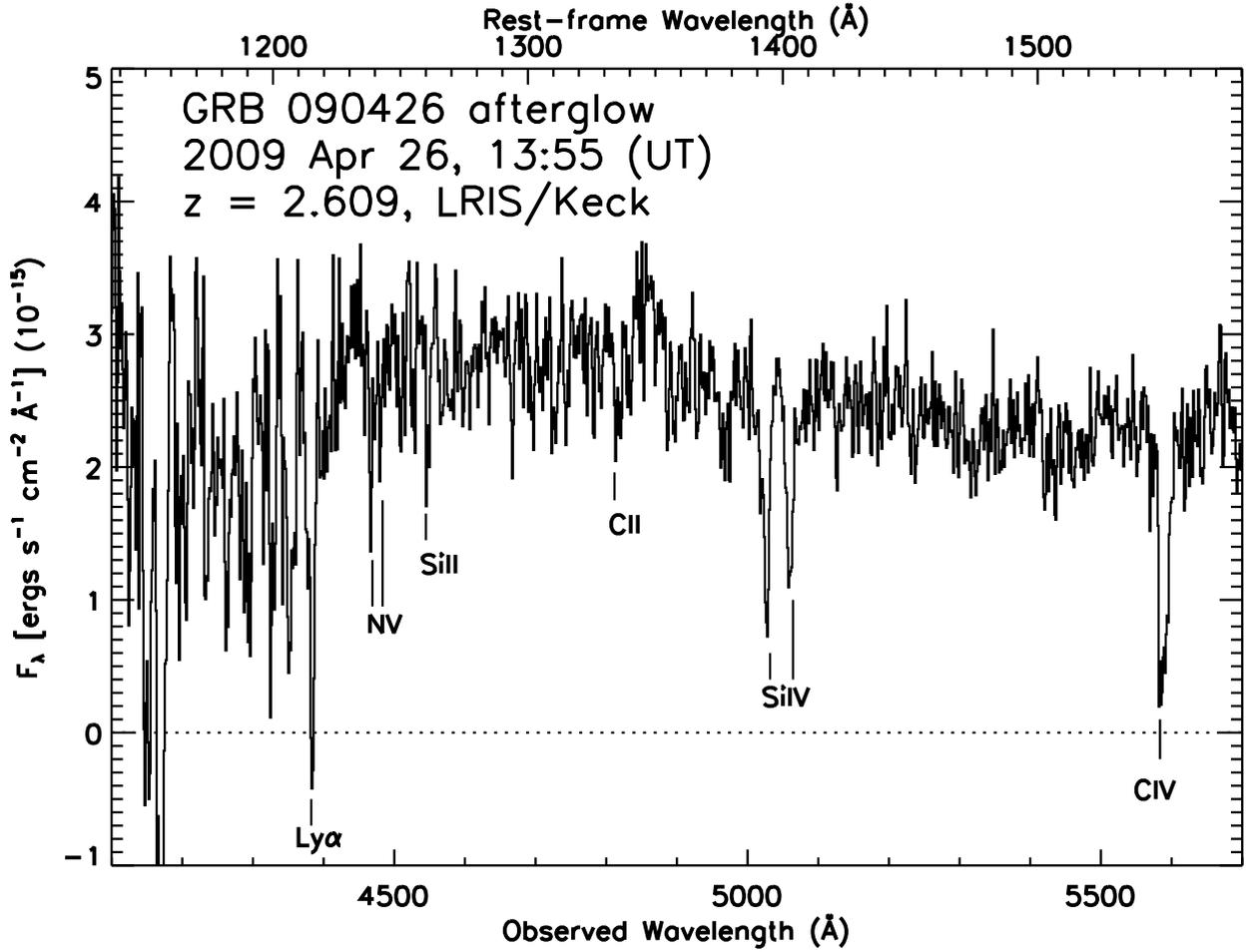}
\caption{Keck spectrum of the GRB 090426 afterglow. The spectrum was
  observed with LRIS on Keck~I at 13:55 on 2009 April 26, $\sim$1.1
  hr after the BAT trigger. The observations were conducted in
  photometric conditions. The data were
  reduced using IRAF, and have been corrected for a heliocentric
  velocity of $-16.88$ km s$^{-1}$. We plot both the observed
  wavelength (lower abscissa) and the rest-frame wavelength at our
  redshift of 2.609 (upper abscissa). We note detections of the
  Ly-$\alpha$, N~V, Si~II, C~II, Si~IV, and C~IV features at this 
  redshift.}
\label{fig:spec}
\end{figure*}

The data were reduced using IRAF\footnote{IRAF is distributed by NOAO,
  which is operated by AURA, Inc., under cooperative agreement with
  the NSF.}. We used the \texttt{lrisbias} IRAF task distributed by the
W.\ M.\ Keck Observatory to subtract overscan from the LRIS images,
and apply a wavelength correction based on our internal lamp
observations. The spectrum was extracted using an optimal extraction
algorithm, with deviant pixels identified and rejected based upon the
assumption of a smoothly varying profile. We flux calibrated the data
using observations of HZ 43 to
derive a sensitivity curve which was then applied to the GRB 090426
afterglow observation. Finally, we corrected for a heliocentric velocity
of $-16.88$ km s$^{-1}$ and corrected the spectrum to rest-frame
wavelengths. Our spectrum is shown in Figure \ref{fig:spec}.

\subsection{Analysis and Interpretation}

In our analysis of the afterglow spectrum, we initially observed a set
of absorption features at 4387~\AA, 5030~\AA, 5061~\AA, and
5592~\AA. We identify these features as Ly-$\alpha$, Si~IV
$\lambda$1394, Si~IV $\lambda$1403, and the blended C~IV
$\lambda\lambda$1548, 1551 doublet at a common redshift of $z = 2.609$. At
this redshift we are also able to identify the N~V
$\lambda\lambda$1239, 1243 doublet, Si~II $\lambda$1260, and C~II
$\lambda$1334 absorption features. We determine the rest-frame
equivalent widths (EWs) for these lines by fitting each line with a
Gaussian using \texttt{splot} in IRAF.

We find that the ionised absorption lines in our spectrum are
saturated, which limits us to determining conservative lower limits
for the column densities of these lines \citep{prochaska06} based on
the relation between EW and column density for saturated lines
\citep{cowie86}. We generally find lower limits for all the saturated columns on the
order of 10$^{14}$ cm$^{-2}$. However, we {\it are} able to calculate
an upper limit for $N_{\rm HI}$ based on the absence of strong
damping wings in the Ly-$\alpha$ absorption feature. From fitting the
line with a Voigt profile, we find an upper limit of $N_{\rm HI} <
3.2 \times 10^{19}$ cm$^{-2}$. Our values for EW and the various column densities are
given in Table~\ref{tab:ews}.

The value of the neutral hydrogen column is very low in comparison to other GRBs: based on the cumulative distribution of
$N_{\rm HI}$ in 28 long-duration GRBs at $z \ge 2$ \citep{chen07}, we
find that the afterglow of GRB 090426 has a lower $N_{\rm HI}$ than
$\sim$90\% of GRB afterglow spectra. Our GRB 090426 afterglow spectrum
also appears to have weaker low-ionisation absorption (Si~II, C~II)
than $\sim$95\% of previous afterglow spectra. This sets GRB 090426
apart as atypical when compared to the host environments of other
GRBs, which generally have much stronger absorption features
\citep{prochaska08b}. Nevertheless, even among ``typical'' long-duration GRBs, such very low
columns are not completely without precedent, and a few long-duration
GRB afterglow spectra are found to have similarly low $N_{\rm H~I}$ to
GRB 090426. Typically, GRB afterglows with Ly-$\alpha$ lines have
column densities of $N_{\rm H~I} \approx 10^{21}$ cm$^{-2}$; one
notable exception is GRB 021004, which has $N_{\rm H~I} \approx 1
\times 10^{19}$ cm$^{-2}$. It is suggested that the low measured
$N_{\rm H~I}$ in that afterglow spectrum is due to ionisation of the H
I by the radiation field of the massive-star progenitor
\citep{fynbo05}.  Another unusual afterglow is associated with the
long/soft GRB 060607; with $N_{\rm H~I} = 6.3 \times 10^{16}$
cm$^{-2}$, it has the lowest H~I column density of any GRB
afterglow. The GRB 060607 spectrum lacks any detection of the N~V
lines, though it does show C~IV and Si~IV absorption at the redshift
of the GRB \citep{prochaska08b}. However, no host galaxy has been
detected for GRB 060607 thus far, down to an $H$-band limiting magnitude of $AB(H) = 26.5$ \citep{Chen09}. By contrast, we do in fact detect the
N~V doublet in the afterglow spectrum of GRB 090426. N~V is thought to originate in the immediate circumburst environment of the GRB, and this absorption feature is
quite typical of most other observed GRB afterglows \citep{prochaska08b}.

Similarly, examples of systems with extremely weak low-ionization
lines, while quite rare, are not unprecedented among ordinary long
bursts: GRB 070125 and GRB 071003 were both found to have extremely
weak host Mg~II absorption systems \citep{cenko08,perley08},
indicative of a low-density galactic environment, possibly in a tidal
tail or halo.

\section{The Host Galaxy of GRB 090426}
\subsection{Imaging}
\begin{figure*}
\centerline{
\includegraphics[scale=0.75,angle=0]{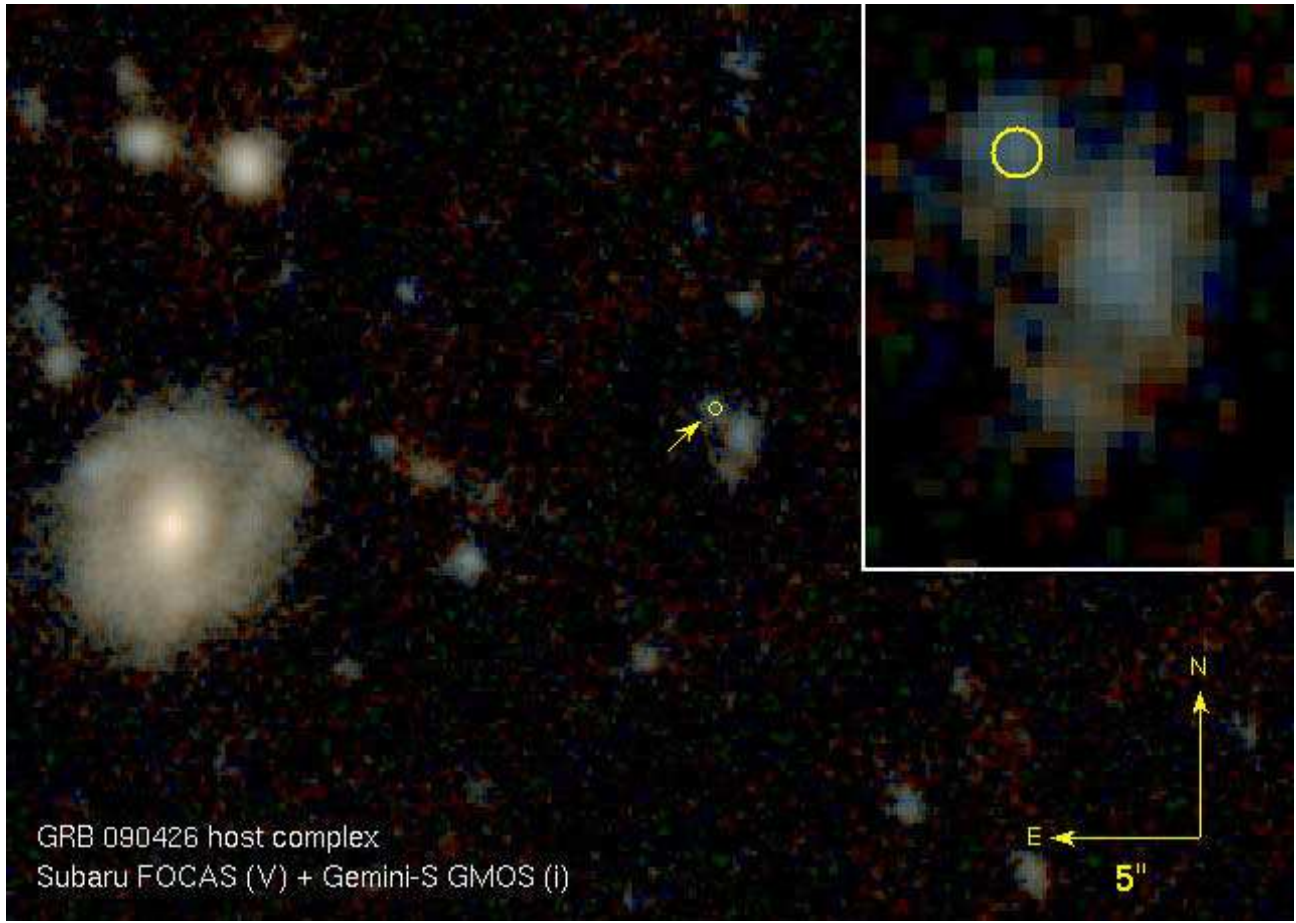}}
\caption{False-colour optical image of the host-galaxy field from
  combined $i$-band data from GMOS-S on Gemini South and $V$-band data
  from FOCAS on Subaru.  A magnified region of the host complex is
  inset at top right.  The afterglow position identified by our LRIS
  acquisition imaging is shown in both images as a yellow circle of
  radius 0.2$\arcsec$ (2$\sigma$) and is consistent with the northeast component
  of the complex. The large galaxy 18\arcsec\ to the East of the host complex is that noted by \citet{davanzo09}.}
\label{fig:imaging}
\end{figure*}

On 2009 May 21 we imaged the field using GMOS-S on Gemini-South
and the $i$-band filter for 20 exposures of 180~s each, for 1~hr
of total integration time.  Images were processed using archival
twilight flatfields and fringe corrected within the \texttt{gemini}
package in IRAF.  The following night (2009 May 22) we acquired
additional imaging in the $V$ band using the FOCAS instrument on Subaru.
A total of 9 images of 300~s each were acquired for a total
integration time of 45 min.  Images were processed using standard
techniques in IRAF.  Both optical images show a bright, extended
object with complicated morphology (a bright, elongated object with
fainter lobes of emission to the NE and S) near the afterglow location
(Figure \ref{fig:imaging}).

Finally, on 2009 May 31 we imaged the field using NIRC on the Keck~I
telescope.  A total of 31 exposures of 1~min (10 coadds $\times$ 6~s)
were acquired in the $K$ band, plus 9 in the $H$ band (also 10
$\times$ 6~s), and 9 in the $J$ band (3 $\times$ 20~s).  Images were
processed and stacked using a modified {\it Python}/{\it pyraf} script
originally written by D.~Kaplan.  No object consistent with the
optical band is detected in any filter.  Based on a calibration to
2MASS standards observed in frames taken later in the night, we place
3$\sigma$ limiting magnitudes on the host-galaxy flux of $J > 23.0$,
$H > 22.1$, and $K > 22.0$ mag (Vega).

To calculate the offset of the afterglow relative to the putative host
galaxy, we aligned both the LRIS acquisition image (taken the night of
the burst) and the Subaru $V$-band observation to reference stars in
the Sloan Digital Sky Survey, giving a position of $\alpha$ =
$12^h36^m18^s.052$, $\delta$ = $+32^\circ59\arcmin09.14\arcsec$ (J2000).
This position places the afterglow within 0.1$\arcsec$ (800 pc in
projection) of the centre of the northeastern lobe of the system that
we subsequently identify as the host galaxy complex.

Aperture photometry of the brightest (central) region of the host as
well as the knot at the afterglow location was performed with IRAF
using a 1\arcsec\ radius.  The resulting photometry, corrected for the
modest Galactic extinction ($E(B-V) = 0.017$\,mag; \citealt{sfd98}),
is presented in Table~\ref{tab:imaging}.  In addition to the spatial
coincidence, the identical colours strongly suggest that the two objects
are physically related.  Interpolating to the flux at 1700~\AA\ (see
\citealt{Reddy+2008}), the photometric magnitude of the northeast
component of the host corresponds to a rest-frame UV luminosity of
approximately 0.7 $L_{*}$, or $\sim2L_{*}$ for the entire host complex,
indicating a luminous host galaxy.

\subsection{Spectroscopy}

We obtained an additional late-time spectrum at the afterglow location
with LRIS on 17 June 2009.  Extrapolating the early-time optical light curve, the afterglow flux
should have faded sufficiently such that any emission would be dominated by
host-galaxy light.  Our observations consisted of two 1500~s exposures.  The blue side was configured with the
600/4000 grism, providing coverage of 3500--5500~\AA\ with a scale of
0.62$\arcsec$ pixel$^{-1}$, and a spectral resolution of $\sim 4$~\AA.
The red side employed the 400/8500 grating with wavelength coverage of
5500--10,000~\AA, a scale of 1.18$\arcsec$ pixel$^{-1}$, and a
spectral resolution of $\sim 7$~\AA; however, we do not discuss the
red-side spectrum here, since we are interested primarily in a
detection of Ly-$\alpha$ emission from the host galaxy.  The long,
1$\arcsec$-wide slit was oriented with a position angle of 41.3$^{\circ}$
to capture both the compact ``knot'' at the afterglow location and the
nearby extended galaxy, while an atmospheric dispersion corrector was
utilised to account for differential refraction \citep{filippenko82}.
We are confident that the slit was at the correct location, because
we detected another object at its expected position along the spatial axis of the spectrogram.

The spectra were reduced in a manner identical to that described in \S
3.1.  We find no sign of any flux, either continuum or narrow emission
lines, at the location of the afterglow or host complex.  At $z =
2.609$, Ly-$\alpha$ would fall at $\lambda_{\mathrm{obs}} = 4389$~\AA.
Using observations of the standard star Feige 34 \citep{msb88,o90,s96}
from earlier in the night, we place a limit on any emission-line flux
at this location of $F < 7 \times 10^{-17}$\,erg\,cm$^{-2}$\,s$^{-1}$
(assuming the line was narrow enough to be unresolved in our spectra).
Using the star-formation rate conversions from \citet{b71} and
\citet{k83}, we therefore place a limit on the unobscured
star-formation rate (SFR) at the location of the afterglow and host of
SFR$_{{\rm Ly-}\alpha} < 4$\,M$_{\odot}$\,yr$^{-1}$.  This value is
significantly less than that derived from the k-corrected rest-frame UV (1500\AA) continuum
emission (neglecting extinction corrections), where SFR$_{\mathrm{UV}} = 14.4 \pm 2.0$\,M$_{\odot}$\,yr$^{-1}$; \citep{Kennicutt98}.
However, we note that SFRs derived from Ly-$\alpha$ emission can often
underestimate the true SFR by over an order of magnitude due to
resonant scattering, dust absorption \citep{mkt+03}, and a strong
dependency on the age of the star-forming population (e.g.,
\citealt{v93}).  Furthermore, the night of these spectroscopic
observations was not photometric (variable, thin cloud cover), and
therefore our flux calibration may be in error (though likely at less
than the 50\% level).

\subsection{Models and Interpretation}
We generated synthetic photometry in our measured filters at $z =
2.609$ using the irregular, Sc/d, Sb/c, and elliptical-galaxy
templates in hyper-Z (\citealt{Bolzonella+2000}; templates originally
from \citealt{CWW1980}). The templates were screened by varying
amounts of host-galaxy dust, both with and without the 2175~\AA\ bump
(assuming an LMC and SMC extinction law, respectively) to compare with
the observed colours.  The elliptical and Sb/c models were immediately
ruled out as incompatible with the blue $i-K$ colour implied by the
NIRC non-detections, as were large amounts of dust extinction in any
case.  The highly starburst-dominated irregular template (plus a small
amount of extinction, $A_V \approx 0.4$ mag) is favoured over the slightly
more evolved Sc/d template, though given the limited photometry
available and the simplified nature of the modelling this conclusion
is less robust.  From this examination, it is clear that the
broad-band photometry indicates a stellar population dominated by
young stars.

\subsection{Associating the Galaxy Complex with GRB 090426}
In the optical afterglow spectrum we find no detections of any intervening absorption systems at a redshift of $z<2.609$. Similarly, in our spectroscopic observations of the putative host complex at the position of the afterglow we see no spectral features that would be consistent with contributions from a foreground system.

In the absence of an emission-line redshift of the complex, it is reasonable to ask what the possibility is that this host association is the result of a chance alignment between GRB 090426 and a foreground system at $z<2.609$ (at $z_{\rm complex}$ much larger than $z_{\rm abs}$ the system would be too bright intrinsically). Examining only the northeastern lobe of the host complex and following the prescription in \citet{Bloom+2002}, we estimate a probability of chance alignment between afterglow and the central region of its host galaxy (using an effective radius of 0.25") to be 0.1\%.  More conseratively, if we instead consider the entire host complex (approximately 1.8" radius), the probability of chance alignment is still low at 4\%. Based on the low likelihood of a chance alignment and a lack of spectroscopic evidence supporting the presence of a foreground system, we conclude that this is indeed very likely the host galaxy (complex) of GRB 090426.

\section{Conclusions}

The small astrometric offset from what appears to be a blue host
galaxy initially seems to be difficult to square with the
inference of a very low column density implied by the absorption
spectrum.  However, it is noteworthy that the upper limit probes only
the neutral gas; the implied UV luminosity from the bright host system
suggests a large ionising radiation field in and around the
Galactic disks which may have ionised a significant fraction of the
neutral gas along the line of sight to the GRB.  Furthermore, both the
detection of N~V and the significant circumburst density implied by
the bright afterglow indicate that the immediate environment of GRB
090426 is not dramatically different from those of long-duration GRBs
in general.

On the other hand, a genuine halo environment seems unlikely.  Some degenerate merger scenarios involve a significant ($>$ 1 Gyr) delay between initial formation of the system and the merger, suggesting that the positional and temporal coincidence of the afterglow with what could be a starburst induced by tidal interaction with the nearby object would be relatively unlikely. In these
scenarios, the progenitor system is also subject to a systemic
velocity ``kick'' during binary evolution that results in significant
linear motion of the system away from its birthsite
(\citealt{fwh99,bsp99}; although see also \citealt{bkb02}).  For
instance, a binary with a 100 km s$^{-1}$ kick perpendicular to our line of sight that takes 1~Gyr to
merge will travel 100~kpc from its birthsite; in our adopted cosmology\footnote{$H_0 = 72$ km s$^{-1}$ Mpc$^{-1}$, $\Omega_m = 0.3$, $\Omega_\Lambda = 0.7$}
at $z = 2.609$ this amounts to an angular distance of
12.8\arcsec, compared to the observed $<0.2$\arcsec offset.

This does not immediately reject the association of this event with
merger-product progenitors; if such progenitors can merge over a range
of timescales (including relatively short ones), the association of
SHBs with active starbursts would be no surprise.  If GRB 090426 is
interpreted as arising from a merger, this event may suggest that SHBs
may very well be akin to Type Ia supernovae (which appear to be
generated by both long and short production channels;
\citealt{2006ApJ...648..868S}). Indeed, many SHBs to date have shown
little to no appreciable offset from their (sometimes blue) host
galaxies \citep{2008MNRAS.385L..10T}. This event also serves as a spectroscopic example
of the high-redshift short-duration GRB population inferred from spatial associationsin \citet{Berger07}.  While the most direct evidence for a degenerate merger remains the detection of a gravity wave signature (see for
example \citealt{Bloom09}), then GRB 090426, at a large redshift with large E$_{\rm iso}$, would suggest (cf. \citealt{Berger07,Berger09})
unfortunately that a significant number of SHBs detected by BATSE and {\it Swift} occur well outside of the Advanced LIGO volume.

The simplest conclusion from the available observations of the afterglow and host galaxy
is that GRB 090426 is more closely linked with the core collapse of a
massive star.  The implications of this association are no less
profound: they indicate that the mechanism that generates gamma rays
is capable of operating on timescales as short as 0.3~s, imposing
strong demands of the central engine; in the most basic collapsar model for GRBs,
the duration timescale is generally assumed to be at least an order of
magnitude longer (see \citealt{Woosley:2006p189}).  While events like GRB 090426 are probably rare (due to
relative volumetric effects), these inferences also cast significant doubt on the
classification of a large population of what would otherwise have been
considered classical short-duration bursts: if this burst had occurred
at a similar redshift to prototypical SHBs 050509B or 050724 (at $z = 0.2$--0.3) it would have fallen unambiguously within the SHB duration distribution.  This also illustrates the insufficiency of
$T_{90}$ alone as a classification criterion, given the 92.8\% likelihood that GRB 090426 is a member of the short/hard phenomenological class. At minimum, we feel
that at low redshift, the search for accompanying supernova emission
--- an unambiguous sign of a genuine massive-star origin --- remains
vital to properly distinguishing among different progenitor scenarios.
 \\
 \\

We wish to thank the support staff of the W. M. Keck Observatory,
Subaru telescope, and Gemini Observatory for
their hospitality and assistance. We also thank to Jeff Silverman for his assistance during our observations. This paper
made use of data from the Gamma-Ray Burst Coordinates Network (GCN)
circulars. E.M.L.'s participation was made possible in part by a Ford
Foundation Predoctoral Fellowship. J.S.B.'s group is supported in part
by Las Cumbres Observatory Global Telescope Network, and 
NASA/{\it Swift} Guest Investigator grant NNG05GF55G. N.R.B. is 
partially supported by US Department of Energy (DOE) SciDAC grant
DE-FC02-06ER41453 and through the GLAST Fellowship Program (NASA Cooperative Agreement: NNG06DO90A). J.X.P. is
partially supported by NASA/{\it Swift} grant NNX07AE94G and an NSF
CAREER grant (AST-0548180). A.V.F. and his group are grateful for
funding from US NSF grant AST--0607485, DOE/SciDAC grant
DE-FC02-06ER41453, the TABASGO Foundation, Gary and Cynthia Bengier,
and the Richard and Rhoda Goldman Fund. M.M. is supported by a research fellowship from the Miller Institute for Basic Research in Science.

\bibliographystyle{mn2e}
\bibliography{ms}

\clearpage

\clearpage

\begin{deluxetable}{lcc}
\tablecaption{Species Detected in the Keck/LRIS GRB 090426 Afterglow Spectrum}
\tablecolumns{3}
\tablewidth{0pc}
\tablehead{\colhead{Species ($\lambda_0$)} &\colhead{EW$_0$(\AA)} &\colhead{$N_X$ (cm$^{-2}$)}
}
\startdata
Ly $\alpha$ (1215.67~\AA) &2.8 $\pm$ 0.1 &$<$ 3.2 $\times$ 10$^{19}$ \\
N~V (1238.82~\AA) &0.7 $\pm$ 0.1 &$>$ 2.8 $\times$ 10$^{14}$\\
N~V (1242.80~\AA) &0.3 $\pm$ 0.1 &$>$ 1.8 $\times$ 10$^{14}$\\
Si~II (1260.42~\AA) &0.6 $\pm$ 0.1 &$>$ 3.8 $\times$ 10$^{13}$ \\
C~II (1334.53~\AA) &0.3 $\pm$ 0.1 &$>$ 1.0 $\times$ 10$^{14}$ \\
Si~IV (1393.75~\AA) &2.2 $\pm$ 0.1 &$>$ 3.2 $\times$ 10$^{14}$ \\
Si~IV (1402.77~\AA) &1.7 $\pm$ 0.1 &$>$ 3.7 $\times$ 10$^{14}$ \\
C~IV (1548.20~\AA/1550.78~\AA) &3.6 $\pm$ 0.1 &$>$ 9.1 $\times$ 10$^{14}$ \\
\enddata
\label{tab:ews}
\tablecomments{EW$_0$ and $\lambda_0$ are given in rest-frame quantities.}
\end{deluxetable}

\begin{deluxetable}{lcccc}
  \tablecaption{Photometry of the GRB\,090426 Host-Galaxy Complex}
  \tablecolumns{5}
  \tablewidth{0pc}
  \tablehead{\colhead{Filter} & \colhead{Date} & \colhead{Telescope/Instrument} &
    \colhead{Extended Host} & \colhead{Compact Knot} \\ & \colhead{(2009 UT)} & & 
    \colhead{(AB Magnitude)} & \colhead{(AB Magnitude)}
    }
  \startdata
  $V$ & May 22.26 & Subaru / FOCAS & $24.21 \pm 0.15$ & $24.73 \pm 0.15$ \\ 
    $i^{\prime}$ & May 21.05 & Gemini South / GMOS & $24.09 \pm 0.15$ & 
    $24.61 \pm 0.18$ \\
  $J$ & May 31.30 & Keck I / NIRC & $> 23.9$ & $> 23.9$ \\
  $H$ & May 31.30 & Keck I / NIRC & $> 23.5$ & $> 23.5$ \\
  $K_{s}$ & May 31.35 & Keck I / NIRC & $> 23.8$ & $> 23.8$ \\
  \enddata
\label{tab:imaging}
\end{deluxetable}

\label{lastpage}

\end{document}